\documentstyle[epsf,12pt]{article}
\begin{document}
\begin{center}
\Large{\bf Tubes, Mono Jets, Squeeze Out and CME.}\\
\large{R.S. Longacre$^a$\\
$^a$Brookhaven National Laboratory, Upton, NY 11973, USA}
\end{center}
 
\begin{abstract}
Glasma Flux Tubes, Mono Jets with squeeze out flow around them plus the Chiral
Magnetic Effect(CME) are physical phenomenon that generate two particle
correlation with respect to the reaction plane in mid-central 20\% to 30\%
Au-Au collision $\sqrt{s_{NN}}$ = 200.0 GeV measured at  RHIC.
\end{abstract}
 
\section{Introduction} 

The flux tube model does well in describing a central Au-Au collision at 
RHIC\cite{QGP,tubevsjet}. However the tubes on the inside of 
colliding central Au-Au will undergo plasma instabilities\cite{PI1,PI2} and
create a locally thermalized system. When one consider less central collisions
the plasma instabilities are reduced and internal tubes begin to add to 
particle correlations described in Ref.\cite{tubevsjet}. In the modelling
of the flux tubes of Ref.\cite{tubevsjet} one finds a connection using 
the model of Ref.\cite{PBMC}. This model is used to describe the surface
flux tubes of the 20\% to 30\% mid-central Au-Au collision $\sqrt{s_{NN}}$ = 
200.0 GeV measured at  RHIC\cite{PBME}. When one consider surface flux tubes
the correlations from these tubes effect particle spectrum above a $p_t$ of
0.8 GeV/c. The surface flux tubes increase with centrality and the
correlations they produce increase along with the number of 
tubes\cite{centrality}. When the whole $p_t$ range is considered the 
correlations of the flux tubes in the whole volume become important. As 
discussed above the plasma instabilities reduce the correlations of the 
internal tubes as seen in the turnover of the correlation with centrality for 
the larger $p_t$ range of Ref.\cite{Daugherity}. Ref.\cite{PBME} used 5 tubes 
to describe the surface for the mid-central 20\% to 30\% Au-Au collision 
$\sqrt{s_{NN}}$ = 200.0 GeV measured at  RHIC. We will add 4 more tubes to 
account for internal tubes in our simulation. In this simulation 6 of the 9 
tubes lie in the reaction plane while the other 3 fluctuate in the regions 
above and below the reaction plane. These in plane tubes drive a strong $v_2$ 
as seen in the two particle correlation of $\Delta\eta$ $\Delta\phi$ shown in 
Figure 1. 

The over lapping region of the mid-central 20\% to 30\% Au-Au collision 
$\sqrt{s_{NN}}$ = 200.0 GeV have two less dense regions above and below the
reaction plane. Partons in these regions can under go hard scattering forming
dijets. When one of the dijets scatter out of the reaction plane its partner 
will head into the region of higher density. This jet will be absorbed locally 
by the dense region and the out going partner will shower into a mono jet.
If one boost along the beam axis to a coordinate system where the out going
parton is perpendicular to the beam, we will be riding along with the comoving
medium where the away parton will deposit its energy. This dense region will
cause a shadowing\cite{shadowing} where soft thermal particles will flow around
this region causing squeeze out\cite{squeeze,ART}. The soft squeeze out 
particles look similar to a Mach shock cone particles spraying about the jet 
particles with an angle of 45$^o$\cite{mach}. 

Topological configurations should occur in the hot Quantum Chromodynamic (QCD) 
vacuum of the Quark-Gluon Plasma (QGP) which can be created in heavy ion 
collisions. These topological configurations form domains of local strong 
parity violation (P-odd domains) in the hot QCD matter through the so-called 
sphaleron transitions. The domains might be detected using the Chiral Magnetic
Effect (CME)\cite{warringa} where the strong external (electrodynamic) magnetic
field at the early stage of a (non-central) collision cause a charge 
separation along the direction of the magnetic field which is perpendicular 
to the reaction plane. Mid-central 20\% to 30\% Au-Au collision $\sqrt{s_{NN}}$
= 200.0 GeV will have a large magnetic field creating such an out of plane 
charge separation. However such out of plane charge separation varies its 
orientation from event to event, either parallel or anti-parallel to the 
magnetic field (sphaleron or antisphaleron). Also the magnetic field can be 
up or down with respect to the reaction plane depending if the ions pass in a 
clockwise or anti-clockwise manner. Any P-odd observable will vanish and only 
the variance of observable may be detected, such as a two particle 
correlation with respect to the reaction plane.

\begin{figure}
\begin{center}
\mbox{
   \epsfysize 7.0in
   \epsfbox{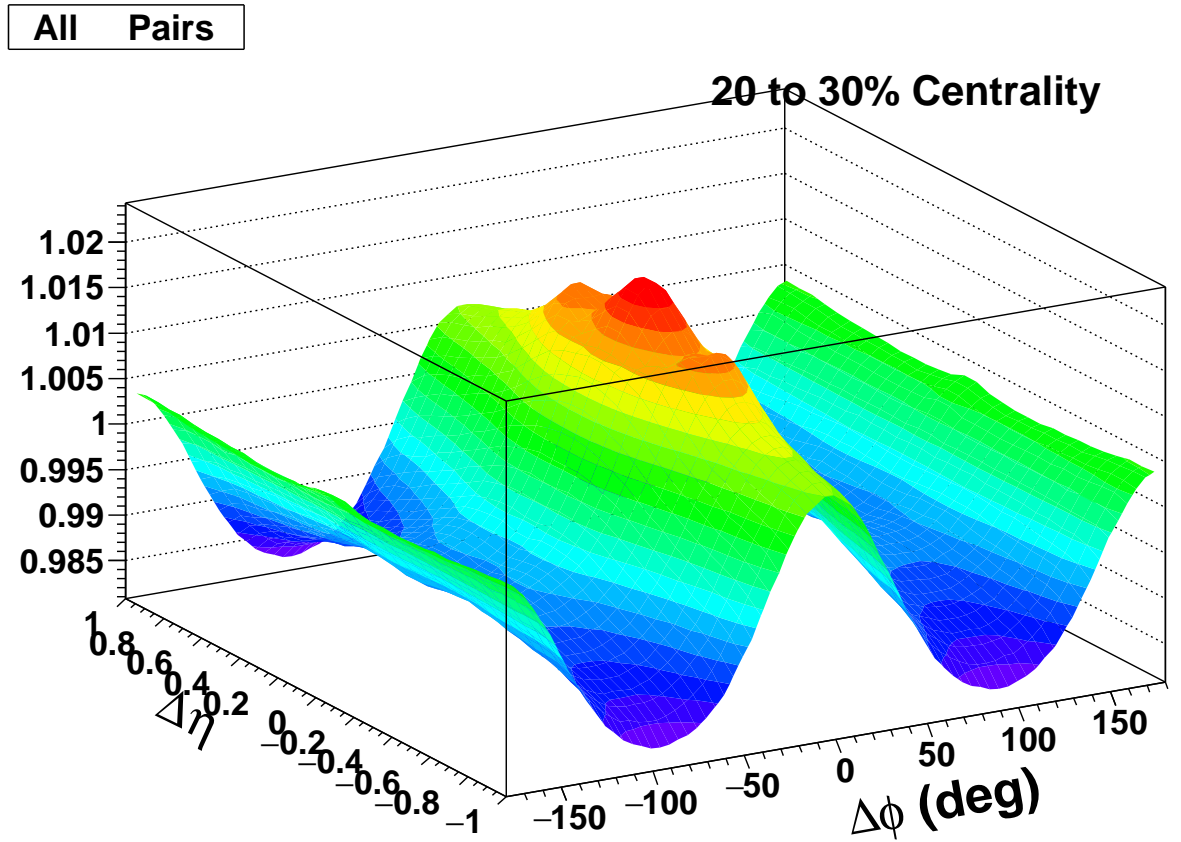}}
\end{center}
\vspace{2pt}
\caption{ A two particle correlation in the 2 dimensional space of 
$\Delta\eta$ $\Delta\phi$. Where $\Delta \eta = |\eta_1 - \eta_2|$  with
$\eta_1$ being the pseudo rapidity of particle 1 and $\eta_2$ being the pseudo 
rapidity of particle 2. Where $\Delta \phi = |\phi_1 - \phi_2|$  with
$\phi_1$ being the azimuthal angles of particle 1 and $\phi_2$ being the 
azimuthal angles of particle 2. All charge particles are generated from the 
model of mid-central 20\% to 30\% Au-Au collision $\sqrt{s_{NN}}$ = 200.0 GeV 
which is described in the text.}
\label{fig1}
\end{figure}

The paper is organized in the following manner:

Sec. 1 is the introduction of all the components present in mid-central 20\% to 
30\%  Au-Au collision $\sqrt{s_{NN}}$ = 200.0 GeV. Sec. 2 presents the two
particle reaction plane dependent correlations we study in this paper. Sec. 3 
further defines the flux tube model and monjets with squeeze out. modifications
of the soft flow particles are added for the squeeze out and the CME particles.
Sec. 4 determines the rate of mono jets from data. Sec. 5 defines the
separation of the reaction plane dependent correlations into two orthogonal 
parts(the in plane and the out of plane). Sec. 6 explores the effect of 
squeeze out. Sec. 7 we show the effects of the Chiral Magnetic Effect(CME) and
calculate the sphaleron mass that we have generated in our simulation. Sec. 8 
presents the summary and discussion.

\section{Angular Correlation with respect to the reaction plane}

There are two very important azimuthal correlations with respect to the 
reaction plane. In mid-central 20\% to 30\% Au-Au collision $\sqrt{s_{NN}}$ 
= 200.0 GeV there is a large magnetic field out of the reaction plane thus 
these two correlations can capture this effect. Also the mono jets of the 
introduction shower out of the reaction plane while the flux tubes lie mainly 
in the plane. The two correlations we will focus on have unique dependence with
pseudo rapidity separation $(\Delta \eta = |\eta_1 - \eta_2|)$. The first 
correlator is given by
\begin{equation}
\langle cos(\phi_1(\eta_1) + \phi_2(\eta_2) - 2\Psi_{RP})\rangle ,
\end{equation}
where $\Psi_{RP}$, $\phi_1$, $\phi_2$ denote the azimuthal angles of the
reaction plane, produced particle 1, and produced particle 2. This two
particle azimuthal correlation measures the sum of $v_1$ at $\eta_1$
of particle 1 with $v_1$ at $\eta_2$ of particle 2. If we would rotate all 
events such 
that $\Psi_{RP}$ = 0.0, then we have
\begin{equation}
\langle cos(\phi_1(\eta_1) + \phi_2(\eta_2))\rangle.
\end{equation}
We also can choose the charge of particle 1 and particle 2 thus have pairs
which have the same sign or like sign and have pairs with opposite sign or 
unlike sign.

The second correlator is given by
\begin{equation}
\langle cos(3\phi_1(\eta_1) - \phi_2(\eta_2) - 2\Psi_{RP})\rangle ,
\end{equation}
where $\Psi_{RP}$, $\phi_1$, $\phi_2$ denote the azimuthal angles of the
reaction plane, produced particle 1, and produced particle 2. This two
particle azimuthal correlation measures the difference of $v_3$ at $\eta_1$
of particle 1 with $v_1$ at $\eta_2$ of particle 2. If we would rotate all 
events such 
that $\Psi_{RP}$ = 0.0, then we have
\begin{equation}
\langle cos(3\phi_1(\eta_1) - \phi_2(\eta_2))\rangle.
\end{equation}
We also can choose the charge of particle 1 and particle 2 thus have pairs
which have the same sign or like sign and have pairs with opposite sign or 
unlike sign.

\section{Flux Tube Model, Mono Jets with squeeze out and the Chiral Magnetic 
Effect(CME)} 

In order to apply the Glasma flux tube model to RHIC mid-central 20\% to 30\% 
Au-Au collision $\sqrt{s_{NN}}$ = 200.0 GeV we use Ref\cite{PBME} to account 
for the surface flux tubes. Ref.\cite{PBME} used 5 tubes to describe the 
surface emissions. To these tubes we add 4 more tubes to account for internal 
tubes in our simulation. In this simulation 6 of the 9 tubes lie in the 
reaction plane while the other 3 fluctuate in the regions above and below the 
reaction plane. These in plane tubes drive a strong $v_2$ as seen in the two 
particle correlation of $\Delta\eta$ $\Delta\phi$ shown in Figure 1. The 
addition of soft particles which approximate the soft particle of a hydro 
system is given by Ref.\cite{hijing} The correlation given by equation 1 is 
referred to as $C_{112}$ in Ref.\cite{monojets}. Jumping forward in figures the 
$C_{112}$ for like sign pairs and unlike sign pairs with and without the CME 
for the pure flux tube model which we describe above is calculated in Figure 
12. One should note the dip in the correlation for like sign pairs. This dip in
the flux tube model is caused by the emissions of aligned quark and
anti-quark pairs coming from gluons in the presence of a strong colored E 
field\cite{PBMC}\cite{central}.   

The single mono jet of the simulation is generated in the over lapping less 
dense regions above or below the reaction plane. Partons in these regions 
can under go hard scattering forming a single dijet. When one of the dijet 
partons scatter out of the reaction plane its partner will head into the 
region of higher density. This jet will be absorbed locally by the dense 
region and the out going partner will shower into the mono jet. If one 
boost along the beam axis to a coordinate system where the out going parton is 
perpendicular to the beam, we will be riding along with the comoving medium 
where the away parton will deposit its energy. This dense region will cause a 
shadowing\cite{shadowing} where soft thermal particles will flow around
this region causing squeeze out\cite{squeeze}. The soft squeeze out particles
look similar to Mach shock cone particles spraying about the jet particles
with an angle of 45$^o$\cite{mach}. 

The mono jets are generated using equation 4 of Ref.\cite{tubevsjet}, where
the the gaussian distributions are charge independent. Normally one might
expect an enhancement of unlike sign pairs. We did not include this in
the mono jet fragmentation since they are emerging out of a medium and are
moving against a strong magnetic field which will force unlike pairs apart(CME).
The angle of the mono jets are spread across either upper or lower azimuthal
range one at a time. The rapidity of the mono jet is also spread across 3 units.
The squeeze out cone is added by adjusting the momentum vector direction of 
some of the soft particles such that we achieve the cone flow pattern around
the mono jet direction. We use a particle by particle each flowing at a
smeared $45^0$ angle to the mono jet axis.
 
The CME is added to the soft particles of the simulation by increasing the
number of like sign pairs moving out of the reaction plane or in the magnetic 
field direction. This is done on a event by event selection. Figure 12 shows 
how the flux tube model C112 correlation splits apart the like sign pairs from 
the unlike sign pairs because of this selection.

\section{Adjusting the level of mono jets in mid-central 20\% to 30\% Au-Au 
collision $\sqrt{s_{NN}}$= 200.0 GeV.}

The correlations of equation 1 and 3 which are called $C_{112}$ and $C_{123}$
in Ref.\cite{monojets} are used to adjust the level of mono jets in our
simulation. The left middle figure of Figure 1 of Ref.\cite{monojets} shows 
the correlation of equation 1 for mid-central 20\% to 30\% Au-Au collision 
$\sqrt{s_{NN}}$= 200.0 GeV. All charge tracks are used in the correlation.
The right middle figure of Figure 2 of Ref.\cite{monojets} shows 
the correlation of equation 3 for mid-central 20\% to 30\% Au-Au collision 
$\sqrt{s_{NN}}$= 200.0 GeV. Again all charge tracks are used in the correlation.
We can achieve Figure 2 of this paper which should be compared Figure 1 and 2 
of Ref.\cite{monojets} by having one mono jet every 7 collisions on average.
This seems reasonable that monjets occur at this level. The crossing of zero
in $C_{123}$ correlation of equation 3 for larger $\Delta \eta$ is caused by
momentum conservation between the flux tubes. For small $\Delta \eta$ 
fragmenting particles which come from the same flux tube drives a positive 
correlation. However when one consider larger $\Delta \eta$ particles, they
will come from different flux tubes which adds an additional negative sign to
the correlation because of momentum conservation between the flux tubes.

With this adjustment we can now predict what the like sign pairs and the 
unlike sign pairs correlations should be for equation 1($C_{112}$) see 
Figure 3. The spitting apart of the correlation is driven by the CME(see Sec. 
6). The results for equation 3($C_{123}$) see Figure 4 where the spitting is 
less important.

\begin{figure}
\begin{center}
\mbox{
   \epsfysize 7.0in
   \epsfbox{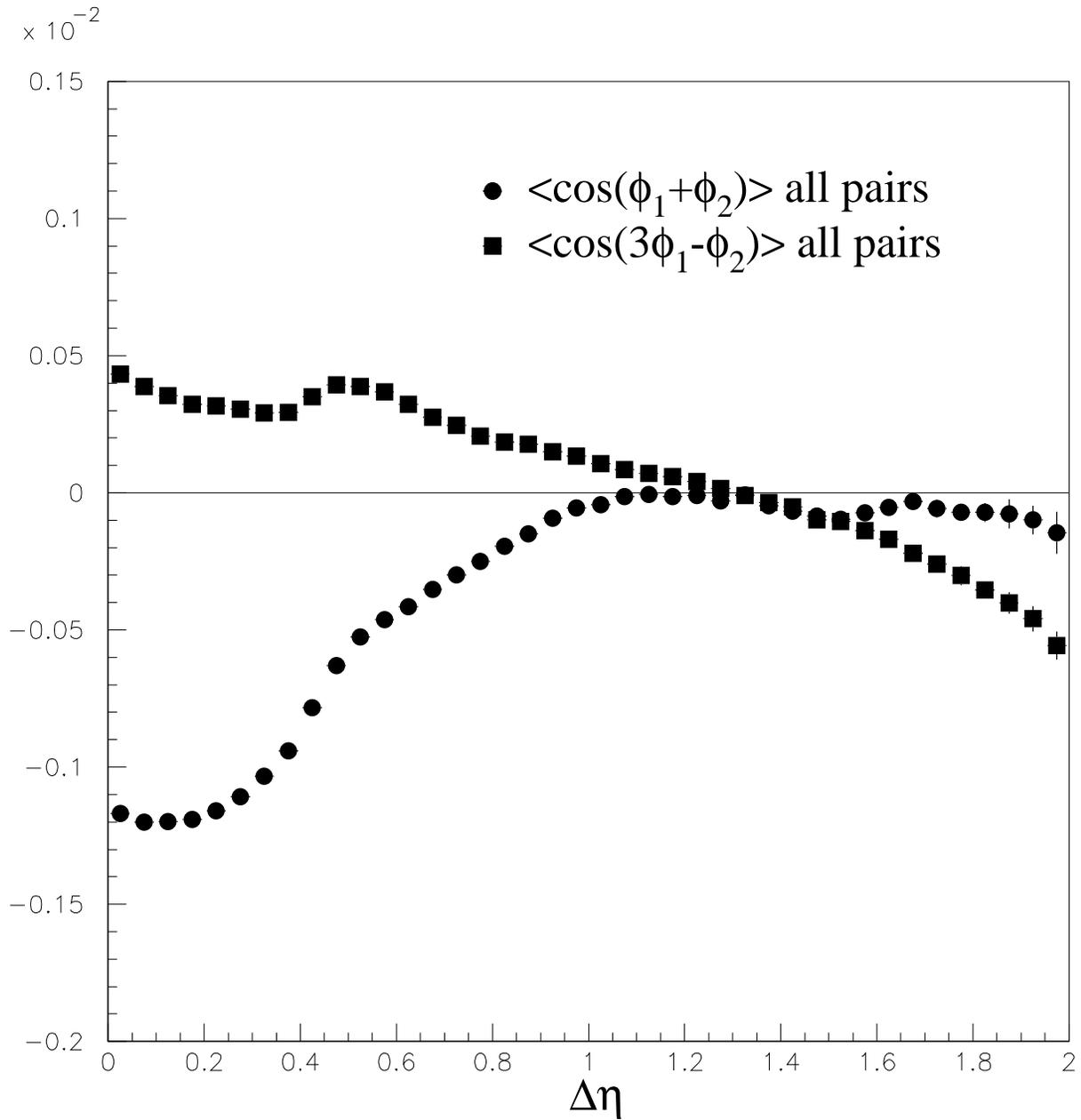}}
\end{center}
\vspace{2pt}
\caption{ Two angular correlations with respect to the reaction plane $C_{112}$
equation 2 $C_{123}$ equation 4. We use equation 2 and 4 since we know the 
reaction plane for each of our generated events. See text for discussion of
the level of mono jets. All charge particles are used in the correlations
and they can be compared to Ref.\cite{monojets}(see text).}
\label{fig2}
\end{figure}
 
\begin{figure}
\begin{center}
\mbox{
   \epsfysize 7.0in
   \epsfbox{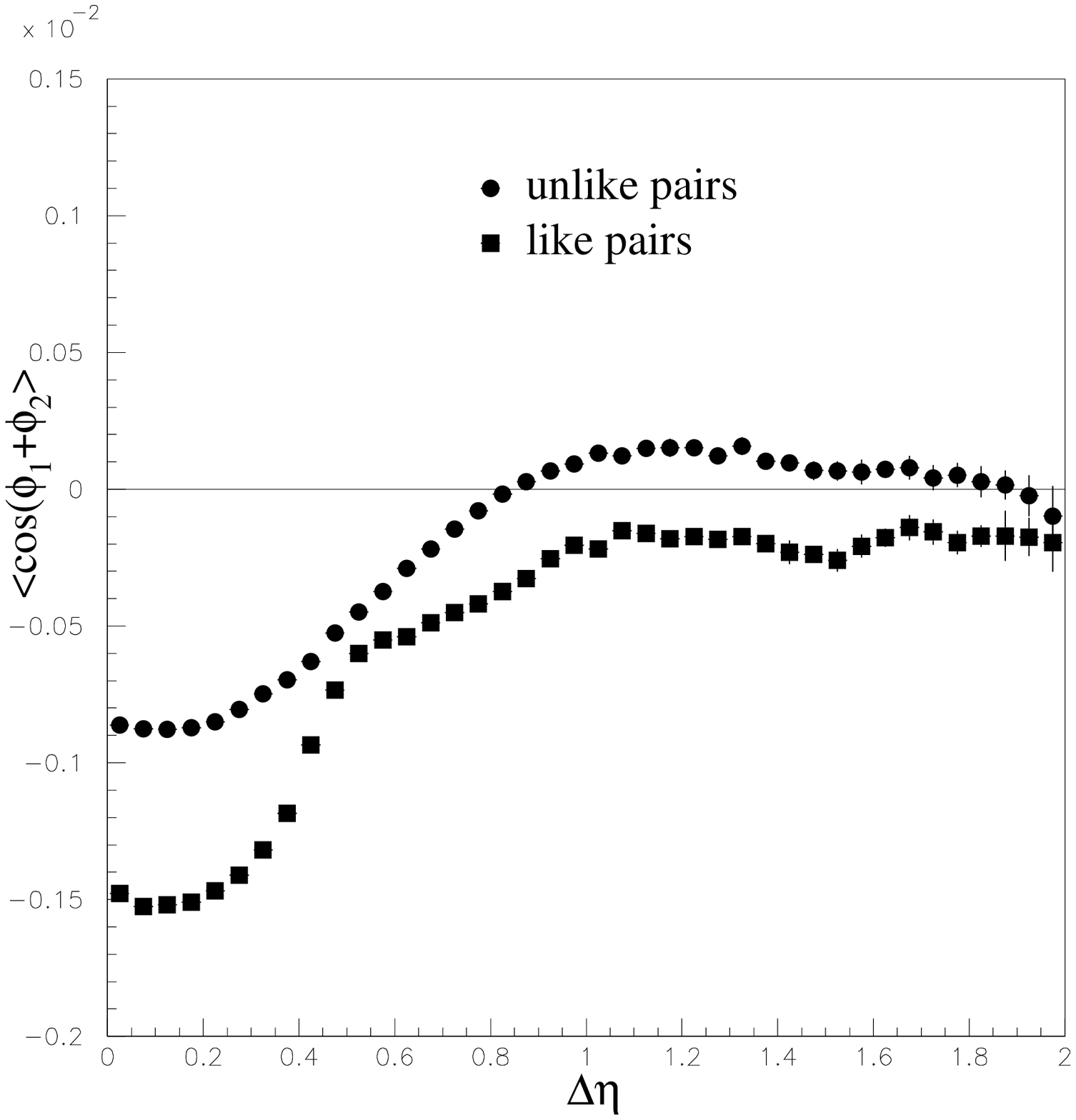}}
\end{center}
\vspace{2pt}
\caption{ Two angular correlations with respect to the reaction plane $C_{112}$
equation 1 for the like sign pairs and the unlike sign pairs. Equation 2 is used
since we know the reaction plane for each of our generated events. See text 
for discussion of the level of mono jets. The spitting apart of the correlation 
is driven by the CME(see Sec. 7).}
\label{fig3}
\end{figure}

\begin{figure}
\begin{center}
\mbox{
   \epsfysize 7.0in
   \epsfbox{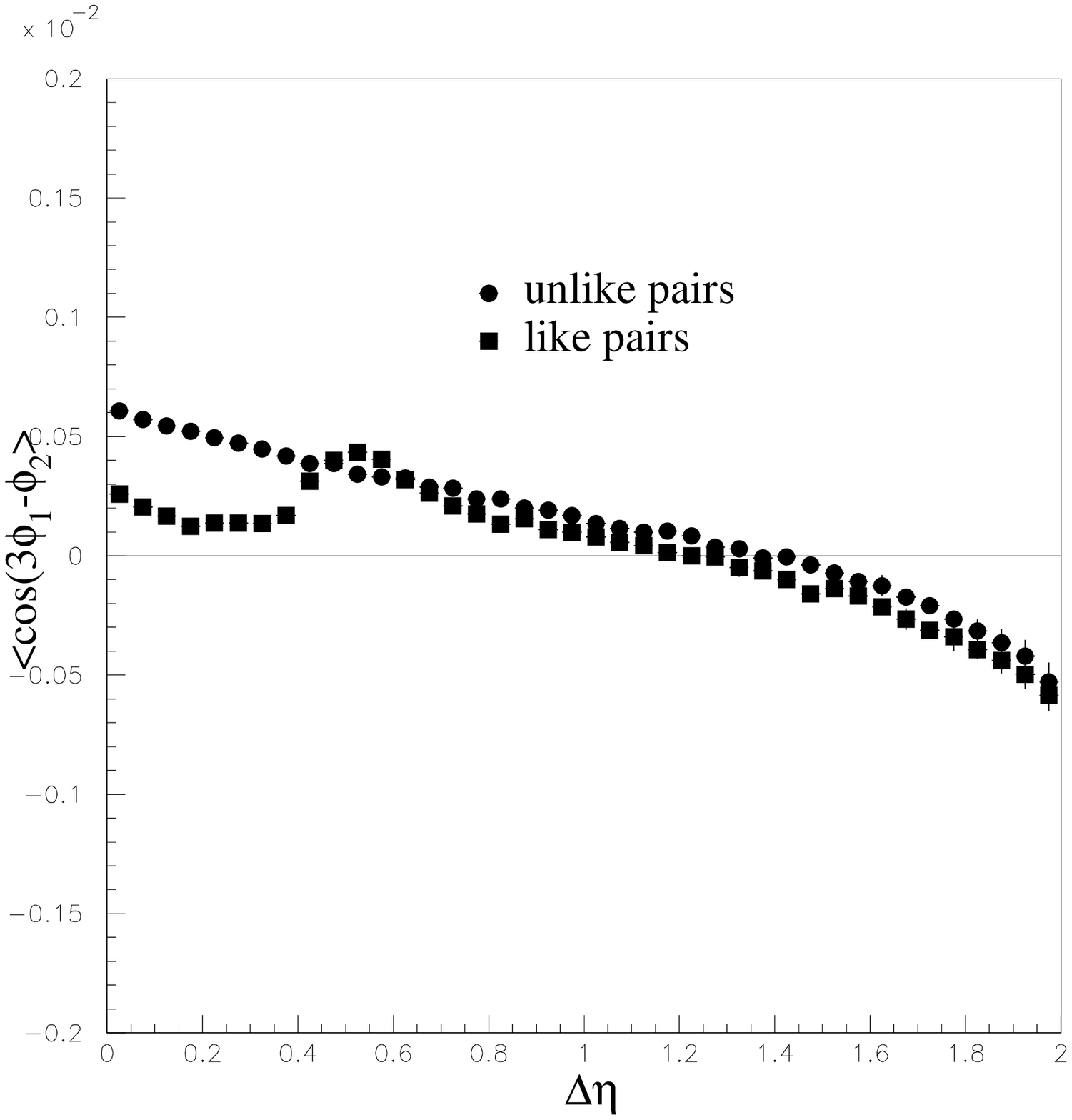}}
\end{center}
\vspace{2pt}
\caption{ Two angular correlations with respect to the reaction plane $C_{123}$
equation 3 for the like sign pairs and the unlike sign  pairs. Equation 4 is 
used since we know the reaction plane for each of our generated events. See 
text for discussion of the level of mono jets. The spitting apart of the 
correlation is driven by the CME(see Sec. 6) and is less important.}
\label{fig4}
\end{figure}

\begin{figure}
\begin{center}
\mbox{
   \epsfysize 7.0in
   \epsfbox{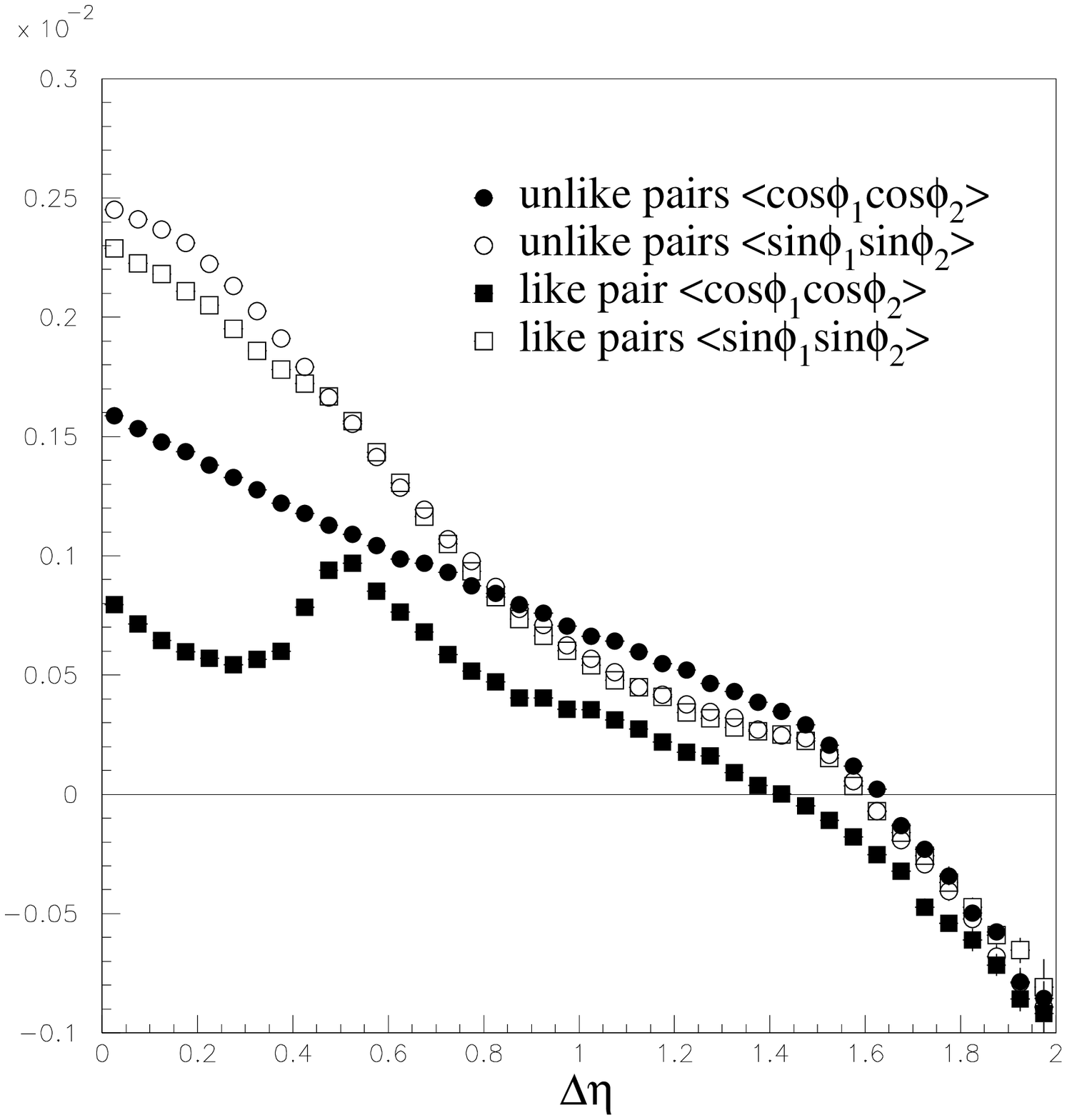}}
\end{center}
\vspace{2pt}
\caption{ An angular correlation $\langle cos(\phi_1(\eta_1) + \phi_2(\eta_2)) 
\rangle$(see Figure 3) for like sign and unlike sign pairs is split up into  
in plane $\langle cos(\phi_1(\eta_1)) cos(\phi_2(\eta_2)) \rangle$ and out of 
plane $ \langle sin(\phi_1(\eta_1)) sin(\phi_2(\eta_2)) \rangle$ parts. The 
in plane is mainly flux tubes while the out of plane is the mono jet effect.
Simulation is for 20\% to 30\% mid-central Au-Au collision $\sqrt{s_{NN}}$ = 
200.0 GeV(see text).}
\label{fig5}
\end{figure}
 
\section{Separation of in plane and out of plane pieces.}

Equation 2 can be separated into an in plane piece and an out of plane piece.
\begin{equation}
\langle cos(\phi_1(\eta_1) + \phi_2(\eta_2)) \rangle = \langle cos(\phi_1(\eta_1)) cos(\phi_2(\eta_2)) \rangle - \langle sin(\phi_1(\eta_1)) sin(\phi_2(\eta_2)) \rangle  .
\end{equation}
Figure 5 shows this separation for our simulation. We see that the cosine 
terms show the correlation driven by the in plane flux tubes while the sine 
terms are driven by the mono jet out of plane effect. Our correlation is for
mid-central 20\% to 30\% Au-Au collision $\sqrt{s_{NN}}$= 200.0 GeV. 
Ref\cite{Csepvsplane} has displayed the same separation for 40\% to 60\% Au-Au 
collision $\sqrt{s_{NN}}$= 200.0 GeV(FIG 8 of Ref\cite{Csepvsplane}). 
We can modify or simulation such that we simulate this centrality. We need to 
reduce our flux tubes by a factor of two along with the over all multiplicity. 
The monjets will only occur once in every thirty events. This separation into
in plane and out of plane is show in Figure 6 and should be compared to 
FIG 8 of Ref\cite{Csepvsplane}. We can also compare the $C_{112}$ the left 
middle figure of Figure 1 of Ref.\cite{monojets} where the mid-central 20\% 
to 30\%  and 40\% to 50\% Au-Au collision $\sqrt{s_{NN}}$= 200.0 GeV are
shown together.  In Figure 7 we show this comparison realizing that we have
to scaled down our 40\% to 50\% by the ratio the multiplicities.

Equation 4 can also be separated into an in plane piece and an out of plane 
piece. 
\begin{equation}
\langle cos(3\phi_1(\eta_1) - \phi_2(\eta_2)) \rangle = \langle cos(3\phi_1(\eta_1)) cos(\phi_2(\eta_2)) \rangle + \langle sin(3\phi_1(\eta_1)) sin(\phi_2(\eta_2)) \rangle  .
\end{equation}
In Figure 8 we show both in plain and out of plane where the in plane
piece drives the correlation through the flux tubes. The out of plane mono jet
gives very little to the correlation and is destroyed by squeeze out(see next
section). 

\begin{figure}
\begin{center}
\mbox{
   \epsfysize 7.0in
   \epsfbox{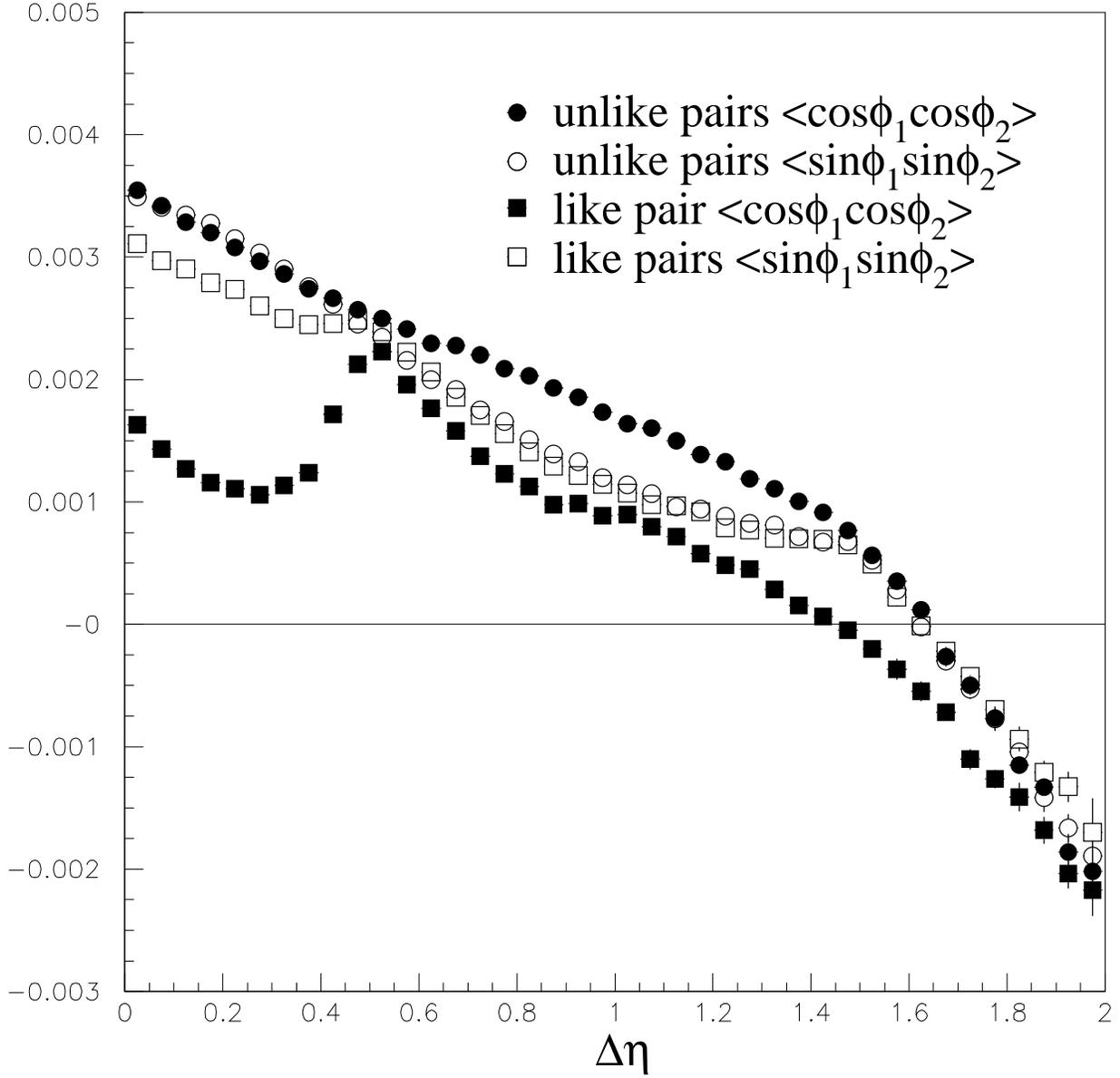}}
\end{center}
\vspace{2pt}
\caption{ An angular correlation $\langle cos(\phi_1(\eta_1) + \phi_2(\eta_2)) 
\rangle$ for like sign and unlike sign pairs is split up into  in plane 
$\langle cos(\phi_1(\eta_1)) cos(\phi_2(\eta_2)) \rangle$ and out of plane 
$ \langle sin(\phi_1(\eta_1)) sin(\phi_2(\eta_2)) \rangle$ parts. The 
in plane is mainly flux tubes while the out of plane is the mono jet effect.
Simulation is for 40\% to 50\% centrality Au-Au collision $\sqrt{s_{NN}}$ = 
200.0 GeV(see text).}
\label{fig6}
\end{figure}

\begin{figure}
\begin{center}
\mbox{
   \epsfysize 7.0in
   \epsfbox{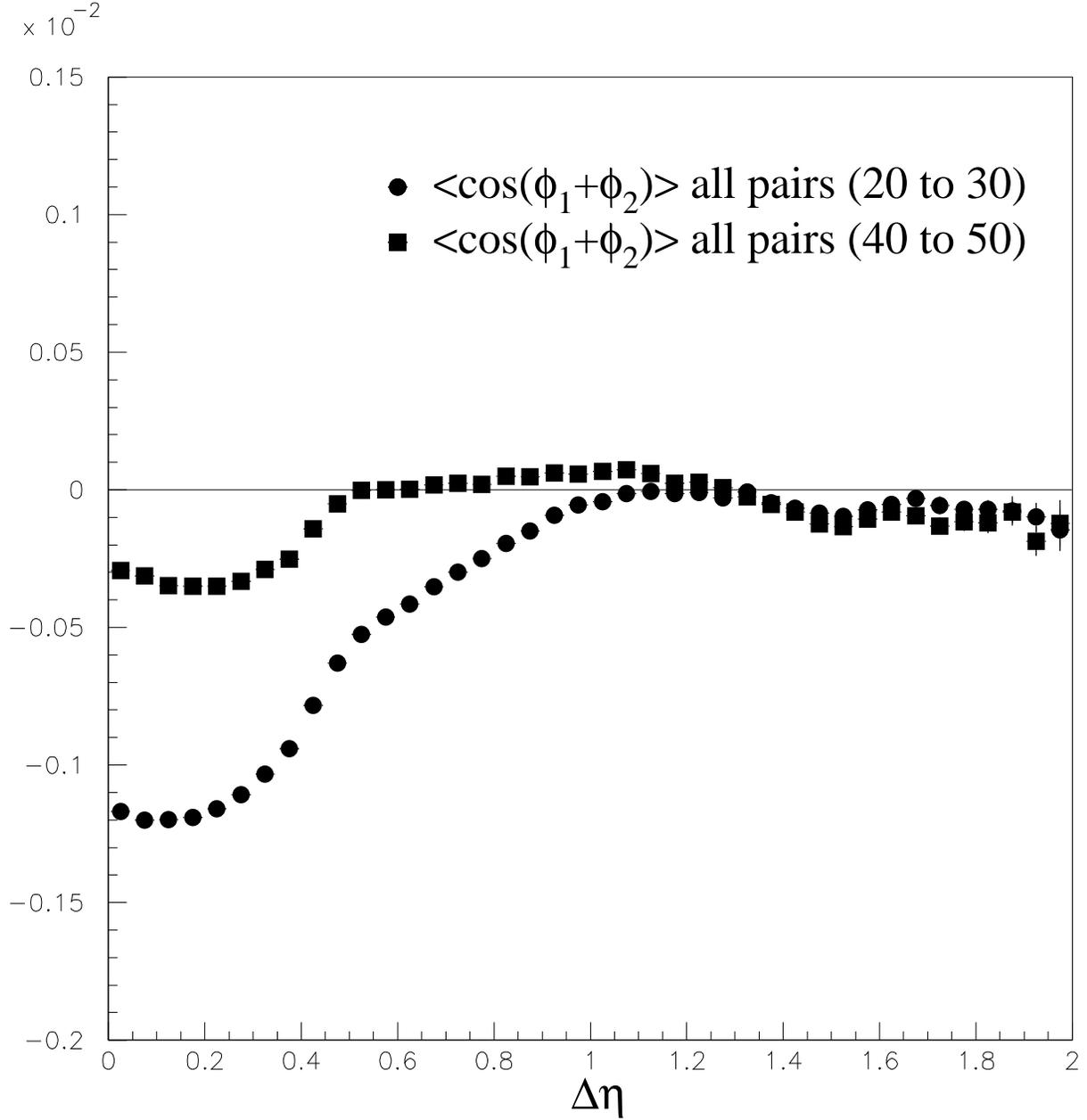}}
\end{center}
\vspace{2pt}
\caption{ The separation into in plane and out of plane is shown in Figure 6 
and should be compared to FIG 8 of Ref\cite{Csepvsplane}. We can also compare 
the $C_{112}$ the left middle figure of Figure 1 of Ref.\cite{monojets} where 
the mid-central 20\% to 30\%  and 40\% to 50\% Au-Au collision $\sqrt{s_{NN}}$
= 200.0 GeV are shown together.  One must realize that we have to scaled down 
our 40\% to 50\% by the ratio the multiplicities.}
\label{fig7}
\end{figure}

\begin{figure}
\begin{center}
\mbox{
   \epsfysize 7.0in
   \epsfbox{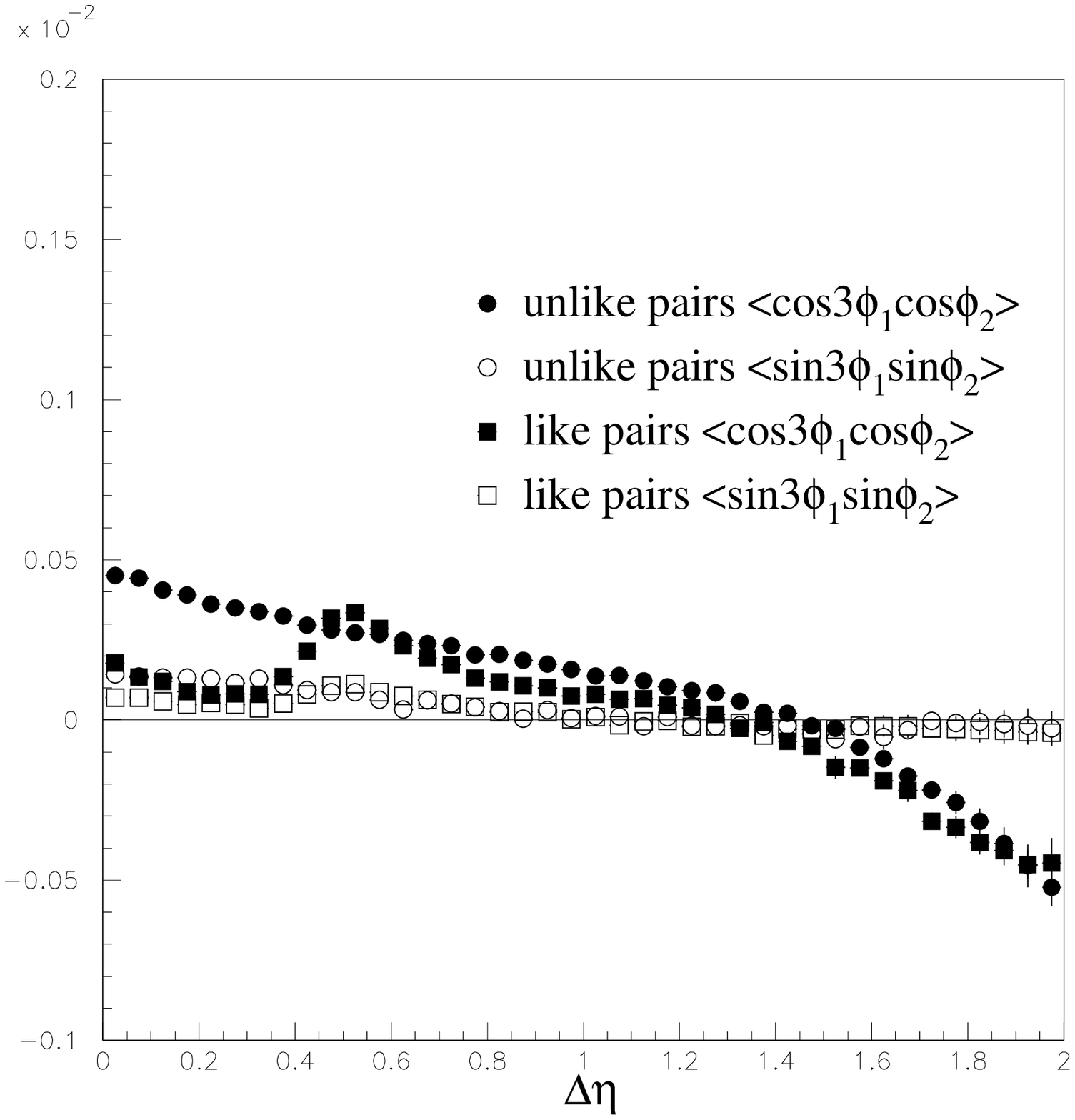}}
\end{center}
\vspace{2pt}
\caption{ Equation 4 can also be separated into an in plane piece and an out of
plane piece. $\langle cos(3\phi_1(\eta_1) - \phi_2(\eta_2)) \rangle = 
\langle cos(3\phi_1(\eta_1)) cos(\phi_2(\eta_2)) \rangle + 
\langle sin(3\phi_1(\eta_1)) sin(\phi_2(\eta_2)) \rangle$. We show both in 
plain and out of plane where the in plane piece drives the correlation through 
the flux tubes. The out of plane mono jet gives very little to the correlation 
and is destroyed by squeeze out(see next section).}
\label{fig8}
\end{figure}

\section{The effect of squeeze out.}

The single mono jet of the simulation is generated in the over lapping less 
dense regions above or below the reaction plane. Partons in these regions 
can under go hard scattering forming a single dijet. When one of the dijet 
partons scatter out of the reaction plane its partner will head into the 
region of higher density. This jet will be absorbed locally by the dense 
region and the out going partner will shower into the mono jet. Again if one 
boost along the beam axis to a coordinate system where the out going parton is 
perpendicular to the beam, we will be riding along with the comoving medium 
where the away parton will deposit its energy. This dense region will cause a 
shadowing\cite{shadowing} where soft thermal particles will flow around
this region causing squeeze out\cite{squeeze}. The soft squeeze out particles
look similar to Mach shock cone particles spraying about the jet particles
with an angle of 45$^o$\cite{mach}. 

In Figure 9 we show the reaction plane dependent correlation equation 2
with and with out squeeze out. Since squeeze out particles flow in the same 
direction as the mono jet they greatly increase the out of plane or negative 
effect on the correlation. We see this clearly if we compare the open points in 
Figure 9(no squeeze out) with the solid points in Figure 12(model without
a mono jet). Where squeeze out particles increase an out of plane effect
for equation 2 it decreases the negative correlation for equation 4 since
the $cos(3\phi_1)$ changes sign as one consider the particles from the mono jet
and particles from the squeeze out differing by 45$^0$. Figure 10 has less 
negative correlation in the mono jet region when squeeze out is present. Without
squeeze out we can achieve the same correlation for $C_{112}$(equation 2) as 
shown in Figure 2 if we have one mono jet per event. Without the squeeze out 
this level of mono jets cause $C_{123}$(equation 4) to be driven negative 
unlike Figure 2(see Figure 11).

\begin{figure}
\begin{center}
\mbox{
   \epsfysize 7.0in
   \epsfbox{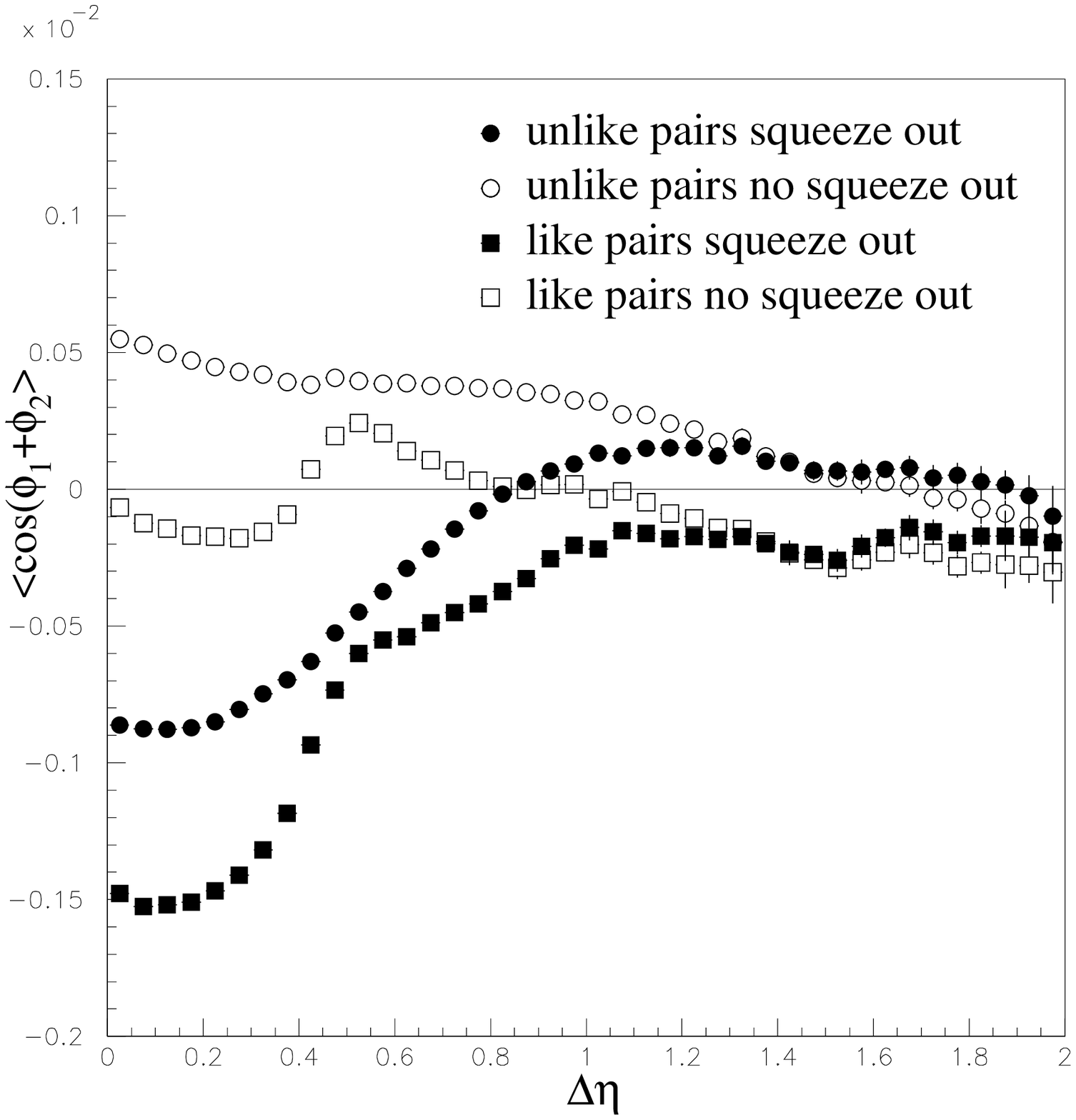}}
\end{center}
\vspace{2pt}
\caption{ Four angular correlations with respect to the reaction plane $C_{112}$
equation 1 for the like sign pairs and the unlike sign  pairs with and without
squeeze out. Equation 2 is used since we know the reaction plane for each of 
our generated events. See text for discussion of squeeze out(Sec. 6).}
\label{fig9}
\end{figure}

\begin{figure}
\begin{center}
\mbox{
   \epsfysize 7.0in
   \epsfbox{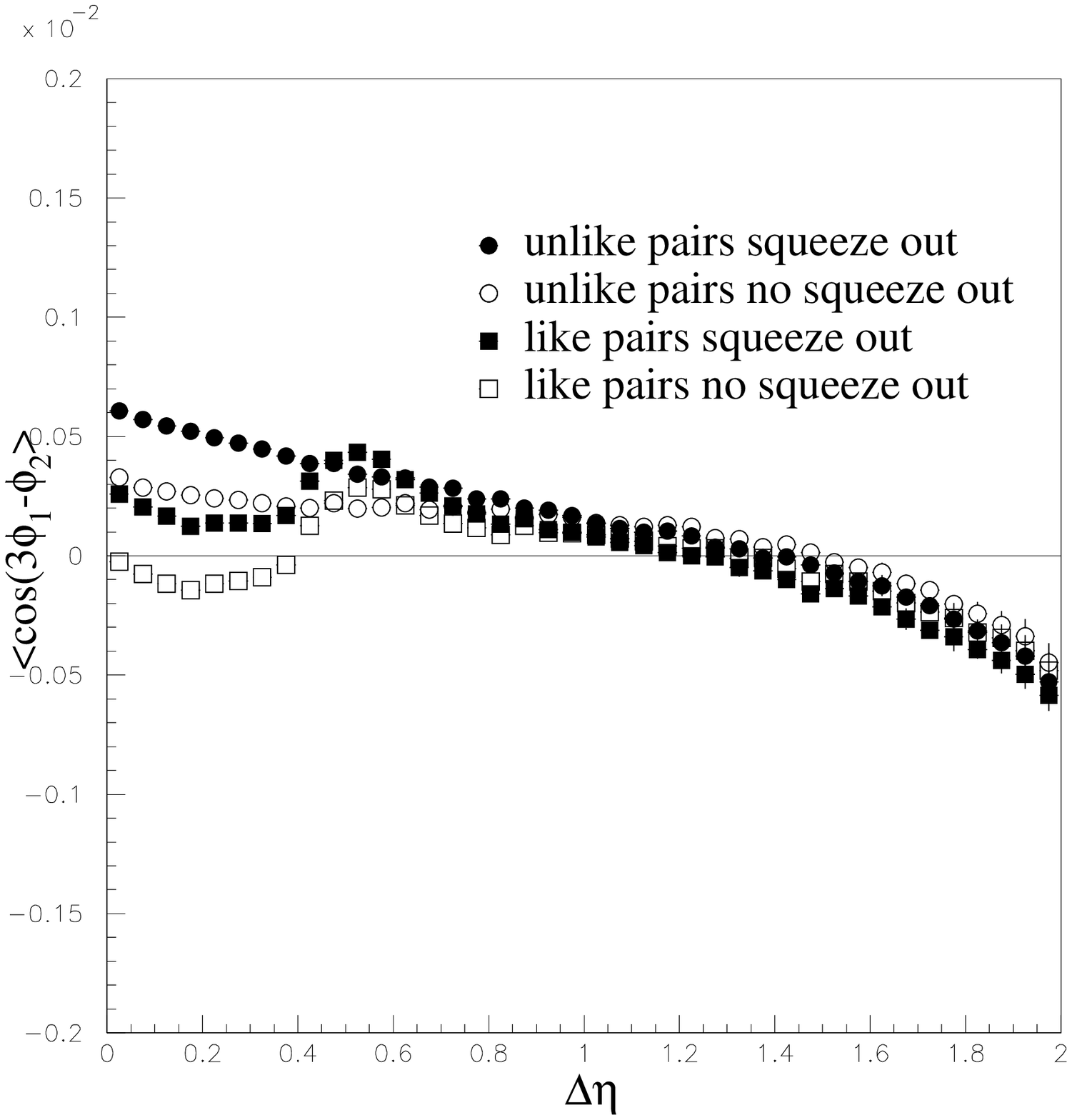}}
\end{center}
\vspace{2pt}
\caption{ Four angular correlations with respect to the reaction plane $C_{123}$
equation 3 for the like sign pairs and the unlike sign  pairs with and without
squeeze out. Equation 4 is used since we know the reaction plane for each of 
our generated events. See text for discussion of squeeze out(Sec. 6).}
\label{fig10}
\end{figure}

\begin{figure}
\begin{center}
\mbox{
   \epsfysize 7.0in
   \epsfbox{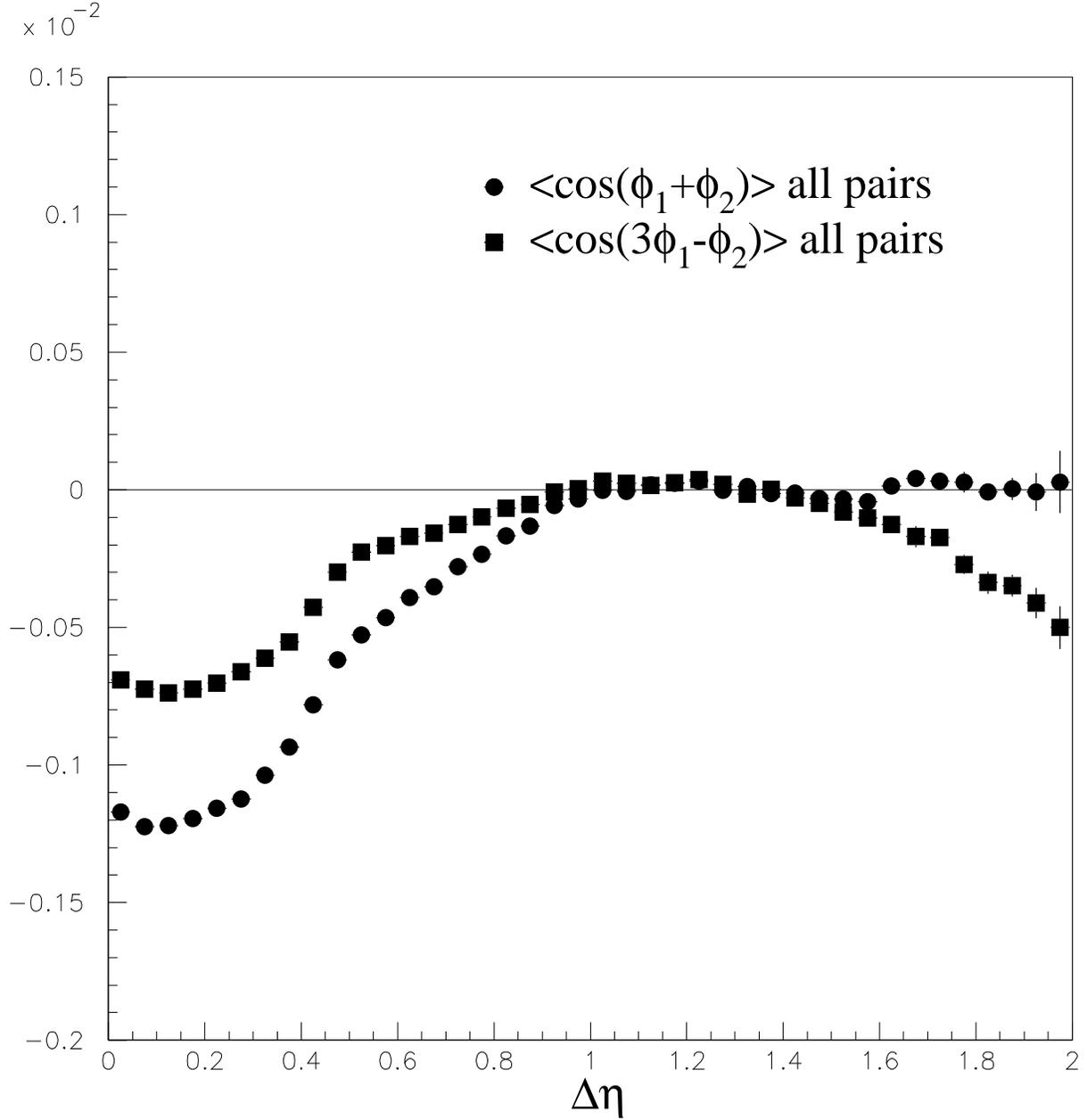}}
\end{center}
\vspace{2pt}
\caption{ Two angular correlations with respect to the reaction plane $C_{112}$
equation 1 and $C_{123}$ equation 3 for all charged pairs. Equation 2 and
equation 4 are used since we know the reaction plane for each of our generated 
events. Without squeeze out we can achieve the same correlation for $C_{112}$
as shown in Figure 2 but without the squeeze out $C_{123}$ is driven negative 
unlike Figure 2.}
\label{fig11}
\end{figure}

\section{The Chiral Magnetic Effect and The Sphaleron Mass}

Topological configurations should occur in the hot Quantum Chromodynamic (QCD) 
vacuum of the Quark-Gluon Plasma (QGP) which can be created in heavy ion 
collisions. These topological configurations form domains of local strong 
parity violation (P-odd domains) in the hot QCD matter through the so-called 
sphaleron transitions. The domains might be detected using the Chiral Magnetic
Effect (CME)\cite{warringa} where the strong external (electrodynamic) magnetic
field at the early stage of a (non-central) collision cause a charge 
separation along the direction of the magnetic field which is perpendicular 
to the reaction plane. Mid-central 20\% to 30\% Au-Au collision $\sqrt{s_{NN}}$
= 200.0 GeV will have a large magnetic field creating such an out of plane 
charge separation. However such out of plane charge separation varies its 
orientation from event to event, either parallel or anti-parallel to the 
magnetic field (sphaleron or antisphaleron). Also the magnetic field can be 
up or down with respect to the reaction plane depending if the ions pass in a 
clockwise or anti-clockwise manner. Any P-odd observable will vanish and only 
the variance of observable may be detected, such as a two particle 
correlation with respect to the reaction plane. The variance correlation we
will use is equation 1 which is a two particle correlation with respect to
the reaction plane($C_{112}$).

In Figure 12 we plot the $C_{112}$ correlation for like sign pairs and unlike 
sign pairs with and without the CME for the pure flux tube model which we 
describe in Sec 2. One should note the dip in the correlation for like sign 
pairs. This dip in the flux tube model is caused by the emissions of aligned 
quark and anti-quark pairs coming from gluons in the presence of a strong 
colored E field\cite{PBMC}\cite{central}. The CME is added to the soft 
particles of the simulation by increasing the number of like sign pairs moving 
out of the reaction plane or in the magnetic field direction. This is done on a 
event by event selection. Figure 12 shows how the flux tube model $C_{112}$ 
correlation splits apart the like sign pairs from the unlike sign pairs because
of this selection. In Figure 13 we plot our predictions for the $C_{112}$ 
correlation with mono jets for like sign pairs and unlike sign pairs with and 
without the CME. To know what the $C_{112}$ would be without the CME one relies
on models since it is not measurable. The CME generated in our model is 
independent of the addition of mono jets, this can be seen in Figure 14 where 
we plot the difference of unlike sign pairs and like sign pairs for the pure 
flux tube model with and without mono jets.
   
The sphaleron is generated in regions of topological charge with a light quark 
helicity unbalance. In the presence of the large external (electrodynamic) 
magnetic field at the early stage of a (non-central) collision charge will 
separation along the direction of the magnetic field which is perpendicular
to the reaction plane. Thus there is a correlation between like sign charge 
pairs. The effective mass of these pairs will be a measure of the sphaleron 
mass. We can plot $C_{112}$ as a function of effective mass for like sign 
pairs and unlike sign pairs with and without the CME. With a CME the 
$C_{112}$ for unlike sign pairs minus like sign pairs is a positive 
correlation. Without a CME this is also positive but not as large. We plot
the difference of this difference with and without in Figure 15. This plot 
gives the effective mass shape of the sphaleron. 

\begin{figure}
\begin{center}
\mbox{
   \epsfysize 7.0in
   \epsfbox{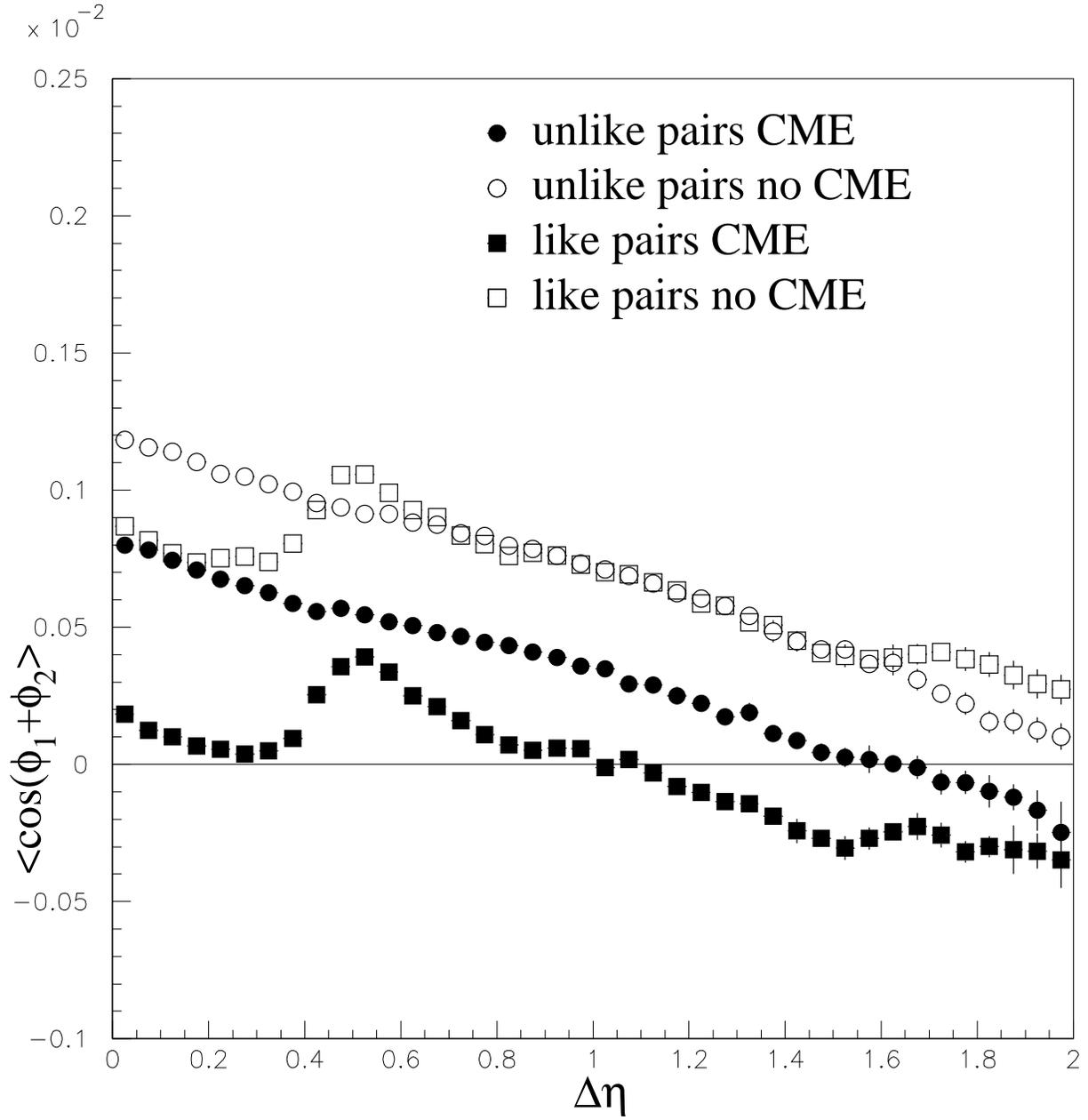}}
\end{center}
\vspace{2pt}
\caption{ Four angular correlations with respect to the reaction plane $C_{112}$
equation 1 for the like sign pairs and the unlike sign pairs for the pure flux 
tube model without mono jets both with and without CME. Equation 2 is used 
since we know the reaction plane for each of our generated events. The spitting 
apart of the correlations is seen when there is a CME.}
\label{fig12}
\end{figure}

\begin{figure}
\begin{center}
\mbox{
   \epsfysize 7.0in
   \epsfbox{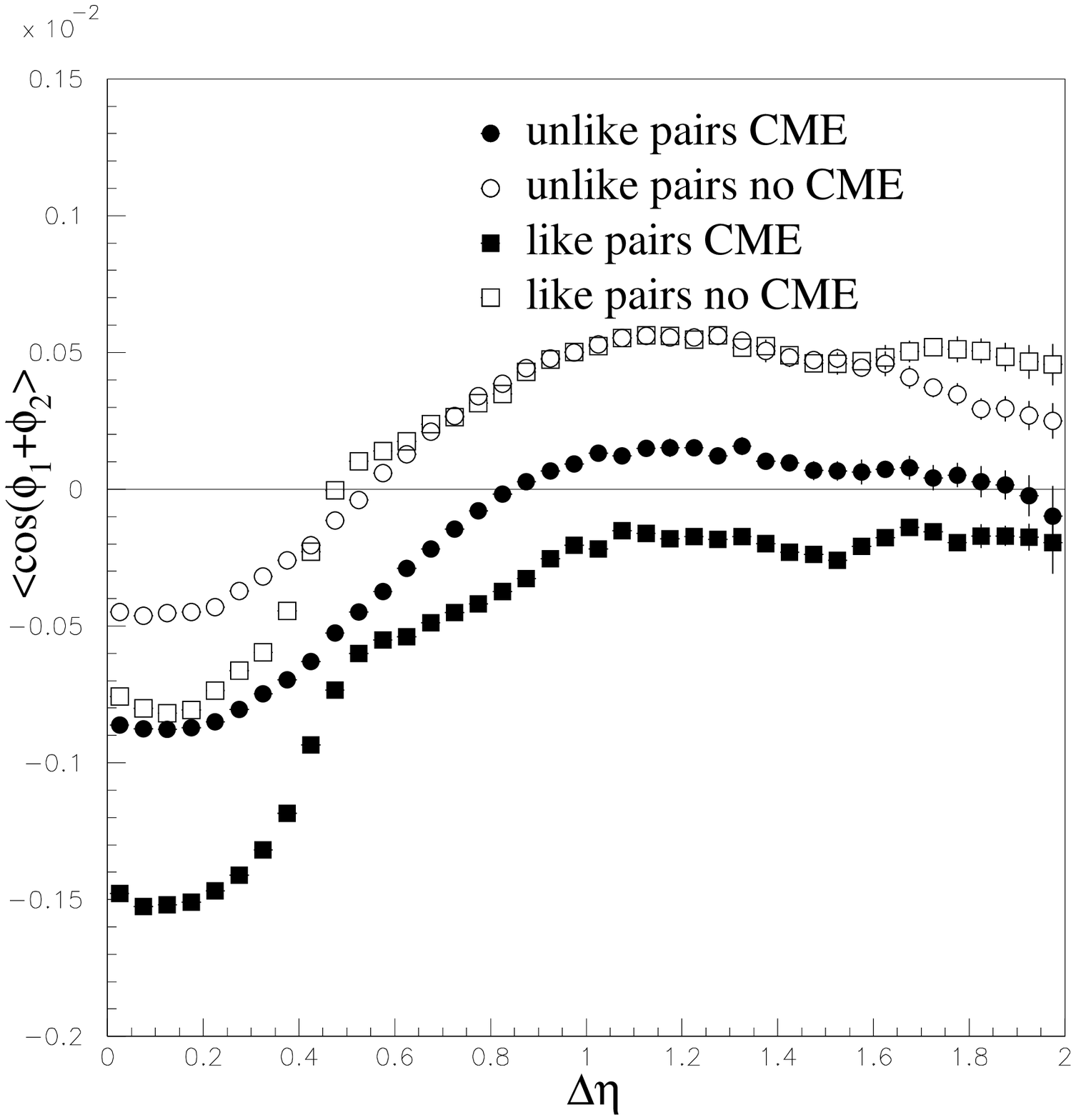}}
\end{center}
\vspace{2pt}
\caption{ Four angular correlations with respect to the reaction plane $C_{112}$
equation 1 for the like sign pairs and the unlike sign pairs for our flux tube
model with mono jets plus squeeze out both with and without the CME. Equation 
2 is used since we know the reaction plane for each of our generated events. 
The spitting apart of the correlations is seen when there is a CME.}
\label{fig13}
\end{figure}

\begin{figure}
\begin{center}
\mbox{
   \epsfysize 7.0in
   \epsfbox{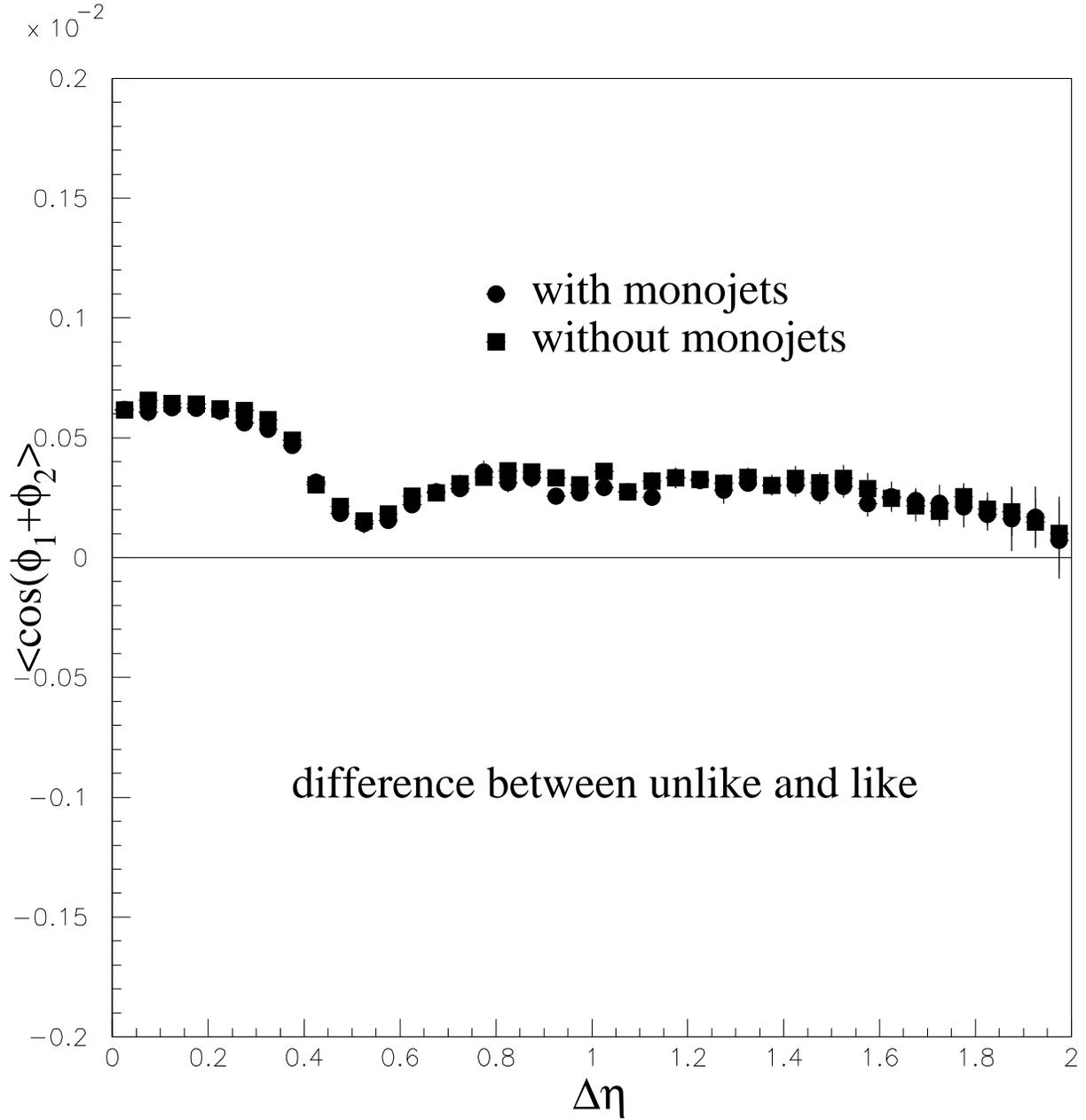}}
\end{center}
\vspace{2pt}
\caption{ Two angular correlations with respect to the reaction plane $C_{112}$
equation 1 for the unlike sign pairs minus the like sign pairs. This difference
is made from Figure 12 the pure flux tube model and Figure 13 our model with
mono jets plus squeeze out both with and without the CME. Equation 2 is used 
since we know the reaction plane for each of our generated events. We see
that the effect of the CME does not depend on the presence of mono jets.}
\label{fig14}
\end{figure}

\begin{figure}
\begin{center}
\mbox{
   \epsfysize 7.0in
   \epsfbox{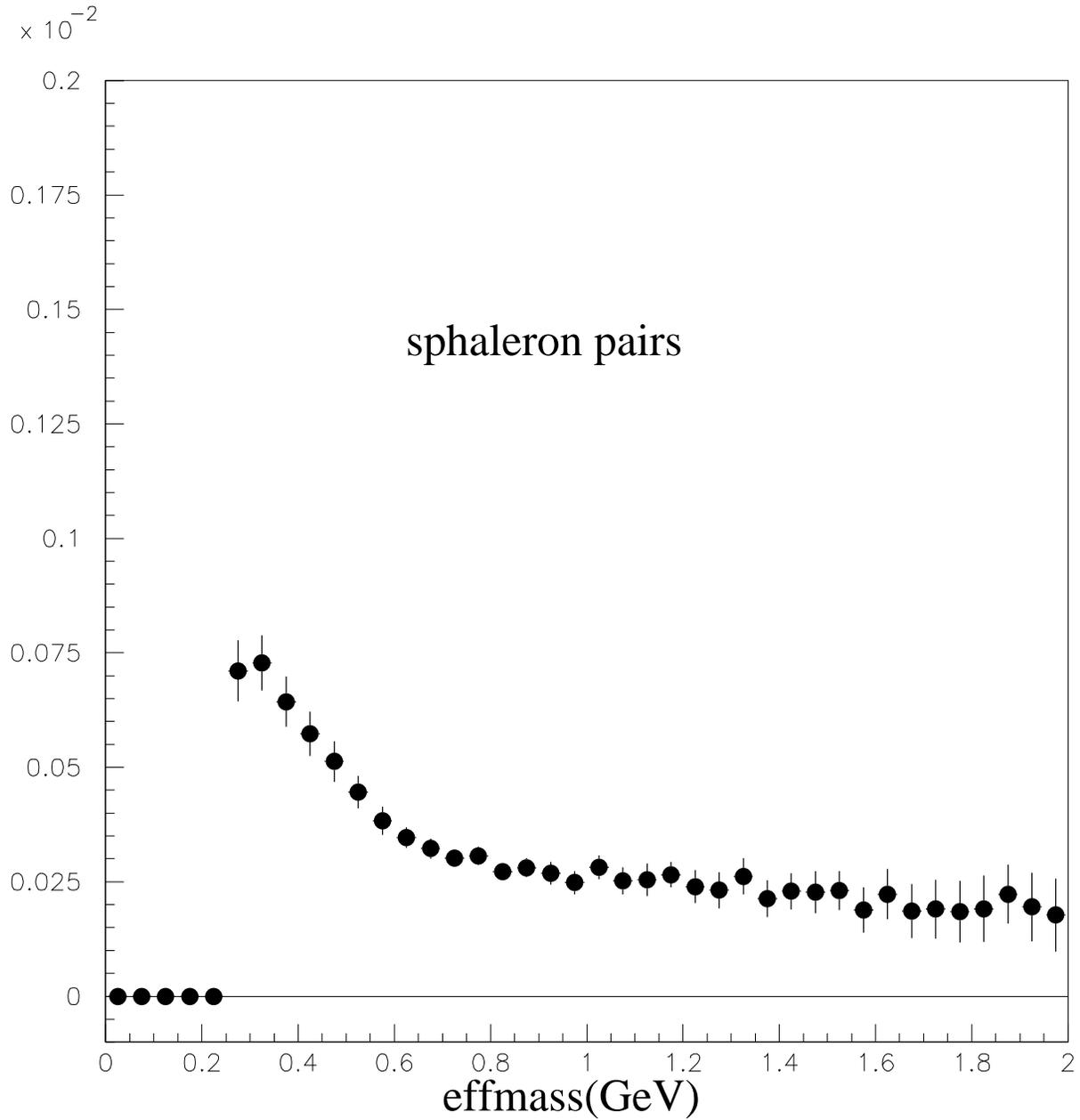}}
\end{center}
\vspace{2pt}
\caption{ The $C_{112}$ as a function of effective mass for unlike sign pairs 
minus like sign pairs with this difference done between with the CME minus
without the CME. The $C_{112}$ for unlike sign pairs minus like sign pairs is 
a positive correlation. Without a CME this is also positive but not as large. 
This difference of difference is shown above and gives the effective mass 
shape of the sphaleron.}
\label{fig15}
\end{figure}

\clearpage

The mass shape of the sphaleron is given by the excess of like sign pairs 
moving out of the reaction plane. The value of this correlation depends
on how many excess pairs per event. The strength of the above correlation 
depends on the magnetic field while the shape depends on the dynamics of the 
sphaleron. So if we would make this difference of difference between two 
systems which are the same except for their magnetic field we would reveal the 
dynamical shape of the sphaleron. Such a controlled study is going to take 
place. $ ^{96}_{44}Ru$  and $ ^{96}_{40}Zr$ are isobars that will collide forming 
systems with a different magnetic field. The charge difference of the two 
colliding systems is 9\%. Thus by forming the $C_{112}$ difference of 
difference between the two systems with $ ^{96}_{44}Ru$ having the bigger 
field being the first term and $ ^{96}_{40}Zr$ being the term subtracted away, 
we will reveal the shape of the sphaleron effect mass and an also prove that 
we are observing the CME.

For completeness in Figure 16 we plot our predictions for the $C_{123}$ 
correlation with mono jets for like sign pairs and unlike sign pairs with and 
without the CME. We see that the CME makes little difference to this 
correlation.

\begin{figure}
\begin{center}
\mbox{
   \epsfysize 7.0in
   \epsfbox{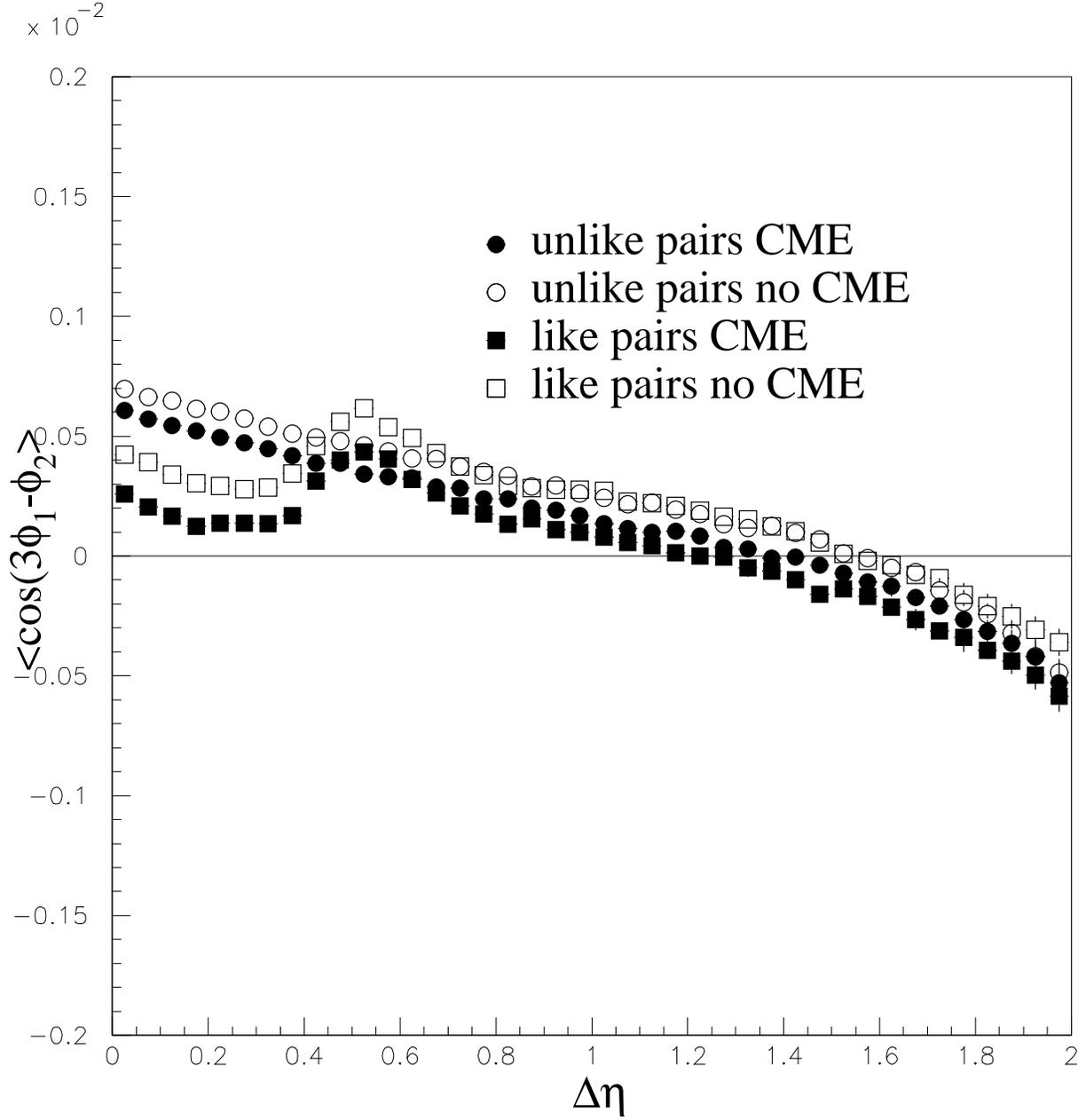}}
\end{center}
\vspace{2pt}
\caption{ four angular correlations with respect to the reaction plane $C_{123}$
equation 3 for the like sign pairs and the unlike sign pairs for our flux tube
model with mono jets plus squeeze out both with and without the CME. Equation 
4 is used since we know the reaction plane for each of our generated events. 
The CME has only a small effect on these correlations.}
\label{fig16}
\end{figure}

\section{Summary and Discussion}

In this paper we explore two very important azimuthal correlations with respect
to the reaction plane. In mid-central 20\% to 30\% Au-Au collision 
$\sqrt{s_{NN}}$ = 200.0 GeV which is the focus of this paper there is a large 
magnetic field out of the reaction plane where these two correlations can 
capture this out of plane effect. Topological configurations form domains of 
local strong parity violation (P-odd domains) in the hot QCD matter through the 
sphaleron transitions. Because of the large magnetic field the domains 
through the Chiral Magnetic Effect (CME)\cite{warringa} will cause charge 
separation along the direction of the magnetic field which is out of plane.
The flux tube model does well in describing Au-Au collision at 
RHIC\cite{QGP,tubevsjet}. See the introduction on how the flux tube model is 
used in this paper. The distribution of flux tubes around the surface of the 
collision generates the large azimuthal flow($v_2$) observed for this 
centrality. The over lapping region of the mid-central Au-Au collision have 
two less dense regions above and below the reaction plane. Partons in these 
regions can under go hard scattering forming dijets. When one of the dijets 
scatter out of the reaction plane its partner will head into the 
region of higher density. This jet will be absorbed locally by the dense 
region and the out going partner will shower into a mono jet. The mono jets 
will shower out of the reaction plane while the flux tubes lie mainly 
in the plane. The two correlations we will focus on have unique dependence with
pseudo rapidity separation $(\Delta \eta = |\eta_1 - \eta_2|)$. The first 
correlator $C_{112}$ is given by
\begin{equation}
\langle cos(\phi_1(\eta_1) + \phi_2(\eta_2) - 2\Psi_{RP})\rangle ,
\end{equation}
where $\Psi_{RP}$, $\phi_1$, $\phi_2$ denote the azimuthal angles of the
reaction plane, produced particle 1, and produced particle 2. This two
particle azimuthal correlation measures the sum of $v_1$ at $\eta_1$
of particle 1 with $v_1$ at $\eta_2$ of particle 2. If we would rotate all 
events such 
that $\Psi_{RP}$ = 0.0, then we have
\begin{equation}
\langle cos(\phi_1(\eta_1) + \phi_2(\eta_2))\rangle.
\end{equation}
We also can choose the charge of particle 1 and particle 2 thus have pairs
which have the same sign or like sign and have pairs with opposite sign sign or 
unlike sign.

The second correlator $C_{123}$ is given by
\begin{equation}
\langle cos(3\phi_1(\eta_1) - \phi_2(\eta_2) - 2\Psi_{RP})\rangle ,
\end{equation}
where $\Psi_{RP}$, $\phi_1$, $\phi_2$ denote the azimuthal angles of the
reaction plane, produced particle 1, and produced particle 2. This two
particle azimuthal correlation measures the difference of $v_3$ at $\eta_1$
of particle 1 with $v_1$ at $\eta_2$ of particle 2. If we would rotate all 
events such 
that $\Psi_{RP}$ = 0.0, then we have
\begin{equation}
\langle cos(3\phi_1(\eta_1) - \phi_2(\eta_2))\rangle.
\end{equation}
We also can choose the charge of particle 1 and particle 2 thus have pairs
which have the same sign or like sign and have pairs with opposite sign or 
unlike sign.

The flux tube model does well in describing our Au-Au collision, however the 
hydro flow of particles quarks and gluons around the flux tube on the fireball 
surface which we will call squeeze out is predicted \cite{shadowing}. 
The single mono jet of the simulation is generated in the over lapping less 
dense regions above or below the reaction plane. Partons in these regions 
can under go hard scattering forming a single dijet. When one of the dijet 
partons scatter out of the reaction plane its partner will head into the 
region of higher density. This jet will be absorbed locally by the dense 
region and the out going partner will shower into the mono jet. If one 
boost along the beam axis to a coordinate system where the out going parton is 
perpendicular to the beam, we will be riding along with the comoving medium 
where the away parton will deposit its energy. This dense region will cause a 
shadowing\cite{shadowing} where soft thermal particles will flow around
this region causing squeeze out\cite{squeeze}. The soft squeeze out particles
look similar to Mach shock cone particles spraying about the jet particles
with an angle of 45$^o$\cite{mach}. The rate of mono jets which gives us 
Figure 2 of this paper which should be compared Figure 1 and 2 
of Ref.\cite{monojets} is one mono jet every 7 collisions on average.
This seems resonable that monjets occur at this level. Figure 3 $C_{112}$ 
and Figure 4 $C_{123}$ are a main perdictions of this paper.

The mass shape of the sphaleron which generates the CME is given by the excess 
of like sign pairs moving out of the reaction plane as function of the 
effective mass of the pairs. The value of this correlation depends on how many 
excess pairs per event. The strength of the above correlation depends on the 
magnetic field while the shape depends on the dynamics of the sphaleron.
So if we would make this difference of difference between two systems which are 
the same except their magnetic field we would reveal the dynamical shape of 
the sphaleron. Such a controlled study is going to take place. $ ^{96}_{44}Ru$  
and $ ^{96}_{40}Zr$ are isobars that will collide forming systems with a
different magnetic field. The charge difference of the two colliding systems \
is 9\%. Thus by forming the $C_{112}$ difference of difference between the two 
systems with  $ ^{96}_{44}Ru$  having the bigger field being the first term 
and $ ^{96}_{40}Zr$ being the term subtracted away, we will reveal the shape 
of the sphaleron effective mass and an also prove that we are observing 
the CME.

\section{Acknowledgments}

This research was supported by the U.S. Department of Energy under Contract No.
DE-AC02-98CH10886.


\begin{thebibliography}{99}
\bibitem{ART} Art Poskanzer points out that squeeze out was first defined as a
system with negatve $v_2$.
\end{thebibliography}
\end{document}